\documentclass[11pt,reqno]{amsart}
\usepackage{amsfonts}
\usepackage{mathrsfs}
\usepackage{amsmath}
\usepackage{amssymb,amsfonts}

\newtheorem{thm}{Theorem}[section]

\newtheorem{coro}{Corollary}[section]\numberwithin{equation}{section}

\def\pf{{\textit {Proof:} }}

\newcommand{\mysection}[1]{\section{#1}\setcounter{equation}{0}}

\newfont{\bb}{msbm10 at 12pt}

\def\S{\hbox{\bb S}}

\def\D{\mathcal D}

\newcommand{\bal}{\begin{aligned}}      \newcommand{\eal}{\end{aligned}}
\newcommand{\ba}{\begin{array}}      \newcommand{\ea}{\end{array}}
\newcommand{\bc}{\begin{center}}     \newcommand{\ec}{\end{center}}
\newcommand{\be}{\begin{enumerate}}  \newcommand{\ee}{\end{enumerate}}
\newcommand{\beq}{\begin{eqnarray}}  \newcommand{\eeq}{\end{eqnarray}}
\newcommand{\beQ}{\begin{eqnarray*}} \newcommand{\eeQ}{\end{eqnarray*}}
\newcommand{\bi}{\begin{itemize}}    \newcommand{\ei}{\end{itemize}}
\newcommand{\bt}{\begin{tabular}}    \newcommand{\et}{\end{tabular}}
\newcommand{\bdm}{\begin{displaymath}} \newcommand{\edm}{\end{displaymath}}

\def\qed{\hfill{Q.E.D.}\smallskip}

\newcommand{\ls}{\setlength{\baselineskip}{12pt}
                 \setlength{\parskip}{3mm}}

\begin{document}

\title[Dirac equation in Kerr-Newman-AdS spacetime]{Nonexistence of time-periodic solutions of the Dirac equation in nonextreme Kerr-Newman-AdS spacetime}

\author{Yaohua Wang}
\address[Wang]{School of Mathematics and Statistics, Henan
University, Kaifeng, Henan 475004, PR China}
\email{wangyaohua@henu.edu.cn}
\author{Xiao Zhang}
\address[Zhang]{Institute of Mathematics, Academy of Mathematics and
Systems Science, Chinese Academy of Sciences, Beijing 100190, PR China}
\email{xzhang@amss.ac.cn}

\date{}

\begin{abstract}

In non-extreme Kerr-Newman-AdS spacetime, we prove that there is no nontrivial Dirac particle which is $L^p$ for $0<p\leq\frac{4}{3}$ with arbitrary eigenvalue $\lambda$, and for $\frac{4}{3}<p\leq\frac{4}{3-2q}$, $0<q<\frac{3}{2}$ with eigenvalue $|\lambda|>q \kappa $, outside and away from the event horizon. By taking $q=\frac{1}{2}$, we show that there is no normalizable massive Dirac particle with mass greater than $\frac{\kappa}{2} $ outside and away from the event horizon in non-extreme Kerr-Newman-AdS spacetime, and they must either disappear into the black hole or escape to infinity, and this recovers the same result of Belgiorno and Cacciatori in the case of $Q=0$ obtained by using spectral methods. Furthermore, we prove that any Dirac particle with eigenvalue $|\lambda|<\frac{\kappa}{2} $ must be $L^2$ outside and away from
the event horizon.

%PACS number: 04.20.Cv

\end{abstract}

\maketitle \pagenumbering{arabic}

\mysection{Introduction}\ls

In \cite{FKSY}, Finster, Kamran, Smoller and Yau proved nonexistence of time-periodic solutions of the Dirac equation in non-extreme Kerr-Newman spacetime, which are normalizable outside and away from the event horizon. The method is to use Chandrasekhar's separation of the Dirac equation \cite{Ch,Ch2}. The results have significant physical implication that a quantum mechanical Dirac particle must either disappear into the black hole or escape to infinity.

However, the situation is more complicated when the cosmological constant is nonzero. In this paper, we study the corresponding problem for non-extreme Kerr-Newman-AdS spacetime which equips with the following metric
\begin{eqnarray*}
\begin{aligned}
\widetilde{g}_{KNAdS}=&-\left[1-\frac{2m{r}-Q^2-P^2}{U} +\kappa^2\big({r}^2+a^2\sin^2{\theta}\big)\right]d{t}^2 \\
& +\frac{V}{U} \sin^2{\theta} d{\varphi}^2+\frac{U}{\Delta_{{r}}}d{r}^2+\frac{U}{\Delta_{{\theta}}}d{\theta}^2 \\
& -2a\sin^2{\theta} \left[\frac{2m{r}-Q^2-P^2}{U}-\kappa^2\big({r}^2+a^2\big)\right]d{t}d{\varphi},
\end{aligned}
\end{eqnarray*}
for some $\kappa >0$, $m>0$, where
\begin{eqnarray*}
\begin{aligned}
\Delta_{{r}}&=\big({r}^2+a^2\big)\big(1+\kappa^2{r}^2\big)-2m{r}+Q^2+P^2,\\
\Delta_{ {\theta}}&=1-\kappa^2a^2\cos^2{\theta},\quad
U={r}^2+a^2\cos^2{\theta},\\
 V&=(2m{r}-Q^2-P^2)a^2 \sin^2{\theta} +U \big({r}^2+a^2\big)\big(1-\kappa^2a^2\big).
\end{aligned}
\end{eqnarray*}
The metric solves the Einstein-Maxwell field equations with an electromagnetic vector potential given by
\begin{eqnarray*}
\begin{aligned}
A=-\frac{Qr}{U}\left(dt-\frac{a\sin^2{\theta}}{\Xi}d{\varphi}\right)
-\frac{P}{U}\cos{\theta}\left(a dt-\frac{r^2+a^2}{\Xi}d\varphi\right)
\end{aligned}
\end{eqnarray*}
where $\Xi=1-\kappa^2a^2$. For our convenience, we re-write the Kerr-Newman-AdS metric as
\begin{eqnarray*}
\begin{aligned}
\widetilde{g}_{KNAdS}=&-\frac{\Delta_{{r}}}{U}\left(dt-\frac{a\sin^2{\theta}}{\Xi}d{\varphi}\right)^2
+\frac{U}{\Delta_{{r}}}d{r}^2 +\frac{U}{\Delta_{{\theta}}}d{\theta}^2\\
&+\frac{\Delta_{{\theta}}\sin^2{\theta}}{U} \left[a dt-\frac{(r^2+a^2)}{\Xi}d{\varphi}\right]^2.
\end{aligned}
\end{eqnarray*}

If $m>0$, $\Delta_{{r}}$ has either four complex roots or two real roots on $r \geq 0$. Throughout the paper, we assume Kerr-Newman-AdS spacetime is non-extreme, i.e., $\Delta_{{r}}$ has two different real roots $0<r_c < r_e$, $r=r_c$ is the Cauchy horizon and $r=r_e$ is the event horizon.
We also denote $M_{r _1}$ the time slice $\{t=constant\}$ of Kerr-Newman-AdS spacetime with $r \geq r_1 >0$.

In this paper, we prove nonexistence of the Dirac equation
\begin{equation}
\left(D+i\lambda \right)\Psi=0, \quad  D=e^\alpha \left(\widetilde{\nabla}_{\alpha}+iA(e_\alpha)\right) \label{Dirac}
\end{equation}
on $r>0$ for real eigenvalue $\lambda$ such that $\Psi$ is $L^p $ on $M_{r _1}$ for $0<p\leq\frac{4}{3}$ with some $r_1 >r_e$,
where $\Psi$ takes the form
\begin{equation}\label{phi}
\Psi=S^{-1}\Phi, \quad
\Phi=e^{-i\left(\omega t+(k+\frac{1}{2})\varphi \right)}
\begin{pmatrix}
X_-(r)Y_-(\theta) \\
X_+(r)Y_+(\theta) \\
X_+(r)Y_-(\theta) \\
X_-(r)Y_+(\theta) \end{pmatrix}
\end{equation}
and $S$ is a diagonal matrix
\begin{equation*}
S=\left|\Delta_{r}\right|^{\frac{1}{4}}
\mbox{diag}
\Big((r+ia\cos\theta)^{\frac{1}{2}},
(r+ia\cos\theta)^{\frac{1}{2}},
(r-ia\cos\theta)^{\frac{1}{2}},
(r-ia\cos\theta)^{\frac{1}{2}}\Big)
\end{equation*}
which vanishes on the event horizon. We also prove nonexistence of eigenspinor $\Psi$ with eigenvalue $|\lambda|>q \kappa $
and $\Psi$ is $L^p $ over $M_{r_1}$ for $0<p\leq\frac{4}{3-2q}$, $0<q<\frac{3}{2}$ with some $r_1 >r_e$.

Following from \cite{FSY, FKSY}, we say that a wave function $\Psi$ is time-periodic with period $T$ if there is a real parameter $\Omega $
such that
\beQ
\Psi(t+T, r, \theta, \varphi)=e^{-i \Omega T} \Psi(t, r, \theta, \varphi).
\eeQ
For time-periodic wave functions, we can separate out the time dependence in a discrete Fourier series
\beQ
\begin{aligned}
\Psi(t, r, \theta, \varphi)= e^{-i \Omega T} & \sum _{n, k \in {\bf Z}}  \sum _{\lambda _{nk} \in {\bf R}}
e^{-i\left(\frac{2n \pi}{T} t+(k+\frac{1}{2})\varphi \right)} \Psi ^{\lambda _{nk}},\\
\Psi ^{\lambda _{nk}} = S^{-1}\Phi ^{\lambda _{nk}}, & \quad
\Phi ^{\lambda _{nk}} =
\begin{pmatrix}
X_- ^{\lambda _{nk}} (r) Y_- ^{\lambda _{nk}} (\theta) \\
X_+ ^{\lambda _{nk}} (r) Y_+ ^{\lambda _{nk}} (\theta) \\
X_+ ^{\lambda _{nk}} (r) Y_- ^{\lambda _{nk}} (\theta) \\
X_- ^{\lambda _{nk}} (r) Y_+ ^{\lambda _{nk}} (\theta) \end{pmatrix}.
\end{aligned}
\eeQ
The Dirac wave function $\Psi$ is normalizable if it is $L^2$ over $M_{r _1}$ for some $r_1 \geq r_e >0$. By taking $q=\frac{1}{2}$ in the above nonexistence of the Dirac equation, it indicates that there is no normalizable massive Dirac particle with mass greater than $\frac{1}{2}\kappa $ outside and away from the event horizon in non-extreme Kerr-Newman-AdS spacetime, and they must either disappear into the black hole or escape to infinity.

Although it contradicts with recent cosmological observations which indicated that our universe should have a positive cosmological constant, the negative cosmological constant as well as the results in the paper may have some physical implications in the theory of strongly coupled superconductors based on the point of view of Anti-de Sitter/Conformal Field Theory correspondence. We refer to the recent introductory overview of relevant theory \cite{CLLY} and references therein.

\mysection{Spin connections on Kerr-Newman-AdS spacetime}\ls

In this section we discuss spin connections for Kerr-Newman-AdS spacetime in terms of Cartan's structure equations.
These results are essentially well-known in the literature. Denote the frame of the Kerr-Newman-AdS metric
\begin{eqnarray*}
\begin{aligned}
e^0&=\sqrt{\frac{\mid\Delta_{{r}}\mid}{U}}\left(dt-\frac{a\sin^2{\theta}}{\Xi}d{\varphi}\right), \ \ \ \ e^1=\sqrt{\frac{U}{\mid\Delta_{{r}}\mid}}d{r}, \\
e^2 &=\sqrt{\frac{U}{\Delta_{{\theta}}}}d{\theta},\ \ \ \
e^3 =\sqrt\frac{\Delta_{{\theta}}}{U}\sin{\theta} \left[a dt-\frac{(r^2+a^2)}{\Xi}d{\varphi}\right]
\end{aligned}
\end{eqnarray*}
and its dual frame
\begin{eqnarray*}
\begin{aligned}
e_0&=\frac{r^2+a^2}{\sqrt{U\mid\Delta_{{r}}\mid}}\left(\partial_t+\frac{a \Xi}{r^2+a^2} \partial_\varphi\right), \ \ \ \ e_1=\sqrt{\frac{\mid\Delta_{{r}}\mid}{U}}\partial_r, \\
e_2&=\sqrt{\frac{\Delta_{\theta}}{U}}\partial_\theta, \ \  \ \
e_3=-\frac{1}{\sqrt {U\Delta_{{\theta}}}}\left(a\sin{\theta} \partial_t+\frac{\Xi}{\sin{\theta}}\partial_\varphi\right).
\end{aligned}
\end{eqnarray*}

It is straightforward to derive the connection 1-form
\begin{eqnarray*}
\begin{aligned}
de^0&=C_{10}^0e^1\wedge e^0+C_{20}^0e^2\wedge e^0+C_{23}^0e^2\wedge e^3, \\
de^1&=C_{12}^1e^1\wedge e^2,\  \ \ \
de^2=C_{12}^2e^1\wedge e^2,\\
de^3&=C_{10}^3e^1\wedge e^0+C_{13}^3e^1\wedge e^3+C_{23}^3e^2\wedge e^3
\end{aligned}
\end{eqnarray*}
where
\begin{eqnarray*}
\begin{aligned}
C_{10}^0&=\partial_r\sqrt{\frac{\mid\Delta_{{r}}\mid}{U }},\ \ \ \ C_{20}^0=-\sqrt{\Delta_{\theta}}\partial_\theta\frac{1}{\sqrt{U }},\\
C_{12}^1&=-\frac{\sqrt{\Delta_{{\theta}}}}{U }\partial_\theta\sqrt{U},\ \ \ \ C_{12}^2=\frac{\sqrt{\mid\Delta_{{r}}\mid}}{U }\partial_r\sqrt{U},\\
C_{10}^3&=-2ar\sqrt{\Delta_{{\theta}}}U^{-\frac{3}{2}}\sin\theta ,\ \
C_{23}^0=2a\sqrt{\mid\Delta_{{r}}\mid}U^{-\frac{3}{2}}\cos\theta ,\\
C_{13}^3&=-\sqrt{\mid\Delta_{{r}}\mid}\partial_r\sqrt{\frac{1}{U}},\ \
C_{23}^3=\frac{1}{\sin\theta }\partial_\theta\bigg(\sqrt{\frac{\Delta_{{\theta}}}{U}}\sin\theta\bigg).
\end{aligned}
\end{eqnarray*}
By Cartan's structure equations, we obtain
\begin{eqnarray*}
\begin{aligned}
\omega^0_{\ 1}&=C_{10}^0e^0-\frac{1}{2}C_{10}^3e^3=-\omega_{01}, \ \
\omega^0_{\ 2}=C_{20}^0e^0+\frac{1}{2}C_{23}^0e^3=-\omega_{02},\\
\omega^1_{\ 2}&=-C_{12}^1e^1-C_{12}^2e^2=\omega_{12}, \ \
\omega^0_{\ 3}=-\frac{1}{2}C_{10}^3e^1-\frac{1}{2}C_{23}^0e^2=-\omega_{03},\\
\omega^1_{\ 3}&=-\frac{1}{2}C_{10}^3e^0-C_{13}^3e^3=\omega_{13}, \ \
\omega^2_{\ 3}=\frac{1}{2}C_{23}^0e^0-C_{23}^3e^3=\omega_{23}.
\end{aligned}
\end{eqnarray*}
Thus spin covariant derivatives take the following forms
\begin{eqnarray*}
\begin{aligned}
\widetilde{\nabla}_{e_0}\Psi=&e_0\Psi-\frac{1}{2}\omega_{01}(e_0)e^0\cdot e^1\cdot\Psi
-\frac{1}{2}\omega_{02}(e_0)e^0\cdot e^2\cdot\Psi\\
&-\frac{1}{2}\omega_{13}(e_0) e^1\cdot e^3\cdot\Psi-\frac{1}{2}\omega_{23}(e_0) e^2\cdot e^3\cdot\Psi,\\
\widetilde{\nabla}_{e_1}\Psi=&e_1\Psi-\frac{1}{2}\omega_{03}(e_1)e^0\cdot e^3\cdot\Psi
-\frac{1}{2}\omega_{12}(e_1)e^1\cdot e^2\cdot\Psi,\\
\widetilde{\nabla}_{e_2}\Psi=&e_2\Psi-\frac{1}{2}\omega_{03}(e_2)e^0\cdot e^3\cdot\Psi
-\frac{1}{2}\omega_{12}(e_2)e^1\cdot e^2\cdot\Psi,\\
\widetilde{\nabla}_{e_3}\Psi=&e_0\Psi-\frac{1}{2}\omega_{01}(e_3)e^0\cdot e^1\cdot\Psi
-\frac{1}{2}\omega_{02}(e_3)e^0\cdot e^2\cdot\Psi\\
&-\frac{1}{2}\omega_{13}(e_3) e^1\cdot e^3\cdot\Psi-\frac{1}{2}\omega_{23}(e_3) e^2\cdot e^3\cdot\Psi.
\end{aligned}
\end{eqnarray*}

From now on, we fix the following Clifford representation
\begin{equation*}
\begin{aligned}
& {e}_0 \mapsto \begin{pmatrix}\ &\ &1 &\ \\ \ &\ & \ &1\\1&\
&\ &\ \\ \ &1&\ &\ \end{pmatrix}, \qquad \,\,\,  {e}_1 \mapsto \begin{pmatrix}\ &\ &-1 &\ \\ \ &\ & \ &1\\1&\ &\ &\ \\ \ &-1&\ &\
\end{pmatrix},\\
& {e}_2 \mapsto \begin{pmatrix}\ &\ &\ &1 \\ \ &\ & 1 &\ \\\ &-1 &\ &\ \\ -1 &\ &\ &\
\end{pmatrix}, \quad {e}_3 \mapsto \begin{pmatrix}\ &\ &\ &i\\ \ &\ & -i &\ \\\ &-i &\ &\ \\ i &\ &\ &\
\end{pmatrix}
\end{aligned}
\end{equation*}
for $\Delta_{{r}}>0$, and fix the following Clifford representation
\begin{equation*}
\begin{aligned}
& {e}_0 \mapsto \begin{pmatrix}\ &\ &-1 &\ \\ \ &\ & \ &1\\1&\ &\ &\ \\ \ &-1&\ &\
\end{pmatrix}, \quad {e}_1 \mapsto \begin{pmatrix}\ &\ &-1 &\ \\ \ &\ & \ &-1\\-1&\
&\ &\ \\ \ &-1&\ &\ \end{pmatrix},\\
& {e}_2 \mapsto \begin{pmatrix}\ &\ &\ &1 \\ \ &\ & 1 &\ \\\ &-1 &\ &\ \\ -1 &\ &\ &\
\end{pmatrix}, \quad {e}_3 \mapsto \begin{pmatrix}\ &\ &\ &i\\ \ &\ & -i &\ \\\ &-i &\ &\ \\ i &\ &\ &\
\end{pmatrix}
\end{aligned}
\end{equation*}
for $\Delta_{{r}}<0$. The spin connections are
\begin{eqnarray*}
\begin{aligned}
\widetilde{\nabla}_{\alpha}\Psi=&e_\alpha\Psi+E_\alpha\cdot\Psi,
\end{aligned}
\end{eqnarray*}
for $\Delta_{{r}}>0$, where
\begin{equation*}
\begin{aligned}
E_0&=-\frac{1}{2}
\begin{pmatrix}
C_{10}^0+\frac{i}{2}C_{23}^0    &  -C_{20}^0-\frac{i}{2}C_{10}^3            &  0 & 0 \\
-C_{20}^0-\frac{i}{2}C_{10}^3   &  -C_{10}^0-\frac{i}{2}C_{23}^0            &  0  &  0\\
0                               &  0                                        & -C_{10}^0+\frac{i}{2}C_{23}^0   &  C_{20}^0-\frac{i}{2}C_{10}^3  \\
0                               &  0                                        &  C_{20}^0-\frac{i}{2}C_{10}^3  &    C_{10}^0-\frac{i}{2}C_{23}^0
\end{pmatrix},\\
E_1&=-\frac{1}{2}
\begin{pmatrix}
0                               &  -C_{12}^1+\frac{i}{2}C_{10}^3 &  0                             &   0 \\
C_{12}^1-\frac{i}{2}C_{10}^3    &      0                         &  0                             &    0\\
0                               &  0                             &  0                             &  -C_{12}^1-\frac{i}{2}C_{10}^3  \\
0                               &  0                             &  C_{12}^1+\frac{i}{2}C_{10}^3  &   0
\end{pmatrix},\\
E_2&=-\frac{1}{2}
\begin{pmatrix}
0                               &  -C_{12}^2+\frac{i}{2}C_{23}^0 &  0                             &   0 \\
C_{12}^2-\frac{i}{2}C_{23}^0    &      0                         &  0                             &    0\\
0                               &  0                             &  0                             &  -C_{12}^2-\frac{i}{2}C_{23}^0  \\
0                               &  0                             &  C_{12}^2+\frac{i}{2}C_{23}^0  &   0
\end{pmatrix},\\
E_3&=-\frac{1}{2}
\begin{pmatrix}
-\frac{1}{2}C_{10}^3-iC_{23}^3  &  -\frac{1}{2}C_{23}^0-iC_{13}^3     &  0  & 0 \\
-\frac{1}{2}C_{23}^0-iC_{13}^3  &  \frac{1}{2}C_{10}^3+iC_{23}^3      &  0  &  0\\
0                               &  0                                  & \frac{1}{2}C_{10}^3-iC_{23}^3   &  \frac{1}{2}C_{23}^0-iC_{13}^3  \\
0                               &  0                                  &  \frac{1}{2}C_{23}^0-iC_{13}^3  &    -\frac{1}{2}C_{10}^3+iC_{23}^3
\end{pmatrix}.
\end{aligned}
\end{equation*}

The spin connections can be derived in a similar way for $\Delta_{{r}}<0$.

\mysection{The Dirac equation and nonexistence}\ls

In this section, we use the method in \cite{FSY,FKSY} to prove certain nonexistence of the Dirac equation on non-extreme Kerr-Newman-AdS spacetime. Using Clifford representations for $\Delta_{r} >0$, the Dirac equation (\ref{Dirac}) reduces to the following equation
\begin{equation*}
\D _1 \phi=\D _2 \phi, \qquad
\phi=
\begin{pmatrix}
X_-(r)Y_-(\theta) \\
X_+(r)Y_+(\theta) \\
X_+(r)Y_-(\theta) \\
X_-(r)Y_+(\theta)
\end{pmatrix}
\end{equation*}
where
\begin{equation*}
\begin{aligned}
\D _1 & =
\begin{pmatrix}
-i\lambda r &  0 &  \sqrt{\Delta_{r}}D_{r+} & 0 \\
0   &     i\lambda r   &  0  &  \sqrt{\Delta_{r}}D_{r-}\\
\sqrt{\Delta_{r}}D_{r-}  &  0  & i\lambda r   &  0  \\
0 &  \sqrt{\Delta_{r}}D_{r+} &  0   &     -i\lambda r\end{pmatrix},\\
\D _2 &=
\begin{pmatrix}
a\lambda\cos\theta &  0 & 0 &  \sqrt{\Delta_{\theta}}L_{\theta+} \\
0   &     -a\lambda\cos\theta  & -\sqrt{\Delta_{\theta}}L_{\theta-} & 0 \\
 0  &  \sqrt{\Delta_{\theta}}L_{\theta+}  & a\lambda\cos\theta   &  0  \\
-\sqrt{\Delta_{\theta}}L_{\theta-} & 0 & 0   &     -a\lambda\cos\theta
\end{pmatrix}
\end{aligned}
\end{equation*}
with
\begin{eqnarray*}
\begin{aligned}
D_{r\pm}=&\frac{\partial}{\partial r}\mp\frac{i}{\Delta_{r}}\left(\omega (r^2+a^2)+Qr+(k+\frac{1}{2})\Xi a\right), \\
L_{\theta\pm}=&\frac{\partial}{\partial \theta}\mp\frac{1}{\Delta_{\theta}}\left(\omega a \sin\theta+\frac{(k+\frac{1}{2})\Xi}{\sin\theta}-P\cot\theta\right)\\
&+\frac{1}{2}
\left(\cot\theta+\frac{\kappa^2a^2\sin\theta\cos\theta}{\Delta_{\theta}}\right).
\end{aligned}
\end{eqnarray*}
Thus there exists a real constant $\epsilon$ such that
\begin{eqnarray}\label{radial1}
\begin{aligned}
\frac{d X_+(r)}{dr}-i\alpha_1 X_+(r)-(i\beta_1+\gamma_1)X_-(r)&=0,\\
\frac{d X_-(r)}{dr}+i\alpha_1 X_-(r)+(i\beta_1-\gamma_1)X_+(r)&=0
\end{aligned}
\end{eqnarray}
where
\begin{eqnarray*}
\begin{aligned}
\alpha_1=&\frac{1}{\Delta_{r}}\left(\omega (r^2+a^2)+Qr+(k+\frac{1}{2})\Xi a\right),\\
\beta_1=&\frac{\lambda r}{\sqrt{\Delta_{r}}},\ \ \gamma_1=\frac{\epsilon}{\sqrt{\Delta_{r}}}.
\end{aligned}
\end{eqnarray*}
Similarly, we could get the radial equation for $\Delta_{{r}}<0$:
\begin{eqnarray}\label{radial2}
\begin{aligned}
\frac{d X_+(r)}{dr}-i\alpha_2 X_+(r)-(i\beta_2+\gamma_2)X_-(r)&=0,\\
\frac{d X_-(r)}{dr}+i\alpha_2 X_-(r)+(i\beta_2-\gamma_2)X_+(r)&=0
\end{aligned}
\end{eqnarray}
where
\begin{eqnarray*}
\begin{aligned}
\alpha_2=&\frac{1}{\Delta_{r}}\left(\omega (r^2+a^2)+Qr+(k+\frac{1}{2})\Xi a\right),\\
\beta_2=&\frac{\lambda r}{\sqrt{-\Delta_{r}}},\ \ \gamma_2=\frac{\epsilon}{\sqrt{-\Delta_{r}}}.
\end{aligned}
\end{eqnarray*}

Same as \cite{FSY, FKSY}, we can show that $X_{+}=0$ or $X_{-}=0$ on the horizons can match the solution inside and outside the horizons.
Let $\Psi$ be the solution of (\ref{Dirac}) taking the form (\ref{phi}).
Denote $\star = e_0\cdot e_1\cdot \in End(\S)$, As $\star ^2=id$, $\S$ can be decomposed as $\S=\S ^+\oplus\S ^-$ where
$$\S ^\pm=\left\{\Phi^{\pm}=\frac{1}{2}(id\mp\star \cdot)\Phi\right\}.$$
We say that $\Phi $ satisfies the local boundary condition on horizons either $\Phi ^+ =0$ or $\Phi ^- =0$ on horizons. Since
\begin{equation*}
e_0\cdot e_1=
\begin{pmatrix}
1 &  0 &  0 & 0 \\
0   &     -1   &  0  &  0\\
0  &  0  & -1  &  0  \\
0 &  0 &  0   &     1
\end{pmatrix},
\end{equation*}
that $X_{\pm}=0$ is equivalent to $\Phi^{\pm}=0$ on horizons.

\begin{thm}\label{thm1}
Let $\Psi$ be the solution of (\ref{Dirac}) taking the form (\ref{phi}) on $r>0$ in non-extreme Kerr-Newman-AdS spacetime.
For $0<p\leq\frac{4}{3}$, there is no nontrivial $\Psi$ for arbitrary $\lambda$ which is $L^p$ over $M_{r_1}$ with some $r_1 >r_e$.
\end{thm}
\pf The equation (\ref{radial1}) implies
\begin{eqnarray}
\begin{aligned}
\frac{\partial \Phi}{\partial r}=E\cdot\Phi, \label{dphi}
\end{aligned}
\end{eqnarray}
where
\begin{equation*}
E=
\begin{pmatrix}
i\alpha _1 &  0 &  i\beta _1 +\gamma _1 & 0 \\
0   &     -i\alpha _1   &  0  &  -i\beta _1 +\gamma _1\\
-i\beta _1 +\gamma _1 &  0  & -i\alpha _1  &  0  \\
0 &  i\beta _1 +\gamma _1 &  0   &     i\alpha _1
\end{pmatrix}.
\end{equation*}
Therefore,
\begin{eqnarray*}
\begin{aligned}
\frac{\partial}{\partial r}|\Phi |^2=&\bar{\Phi}^{t}\left(E+\bar{E}^{t}\right)\Phi\\
=&
2\bar{\Phi}^{t}
\begin{pmatrix}
0 &  0 &  i\beta _1 +\gamma _1 & 0 \\
0   &     0   &  0  &  -i\beta _1 +\gamma _1\\
-i\beta _1 +\gamma _1 &  0  &  0  &  0  \\
0 &  i\beta _1 +\gamma _1 &  0   &     0
\end{pmatrix}\Phi.
\end{aligned}
\end{eqnarray*}
So there exist some positive constants $C_1$, $r_1$ such that, on $r_e<r\leq r_1$,
\beQ
\left | \frac{\partial}{\partial r}|\Phi |^2 \right |\leq 2C_1(r-r_e)^{\beta} |\Phi | ^2
\eeQ
for some $\beta>-1$. Thus, outside the zero set of $\Phi$,
\begin{eqnarray*}
| \Phi (r)| \leq |\Phi (s)| \exp  \left(C_1\int_{s} ^{r} (\bar r -r_e)^\beta d \bar r \right).
\end{eqnarray*}
By taking $r\rightarrow r_e$, we conclude $|\Phi |<\infty$ on $r=r_e$. Moreover, taking $s\rightarrow r_e$, we find $|\Phi |=0$ on $r_e <r \leq r_1$ if $|\Phi |=0$ on $r=r_e$. On the other hand, by (\ref{radial1}), we have
\begin{eqnarray*}
\begin{aligned}
\frac{1}{2}\frac{d}{dr}\left(\left | X_+ \right | ^2 - \left | X_-\right | ^2\right)
=&\Re\left(\frac{d X_+}{dr} \bar X_+ \right)- \Re\left(\frac{d X_-}{dr} \bar X _- \right)\\
=&\Re\left(\left(i\alpha_1 X_+ +(i\beta_1+\gamma_1)X_-\right)\bar X_+ \right)\\
&- \Re\left(\left(-i\alpha_1 X_- +(-i\beta_1+\gamma_1)X _+\right) \bar X _-\right)\\
=&\Re\left(i\alpha_1 \left(X_+\bar X_+ + X_-\bar X_-\right)\right)\\
&+\Re\left(i\beta_1 \left( X_-\bar X_+ + X_+\bar X_-\right)\right)\\
&+\Re\left(\gamma_1 \left( X_-\bar X_+ - X_+\bar X_-\right)\right)=0.
\end{aligned}
\end{eqnarray*}
Therefore we obtain
\begin{eqnarray*}
\begin{aligned}
\left| X_+\right| ^2-\left | X_-\right |^2 = C
\end{aligned}
\end{eqnarray*}
for some constant $C$. If $C\neq0$, we can assume $C > 0$ without loss of generality. Thus
\begin{eqnarray*}
\begin{aligned}
\int_{M_{r_1}}\left |\Psi \right | ^p d\mu
>C'\int_{r_1}^\infty r^{1-\frac{3p}{2}}dr =\infty
\end{aligned}
\end{eqnarray*}
for $0<p\leq\frac{4}{3}$.
This contradicts the assumption that $\Psi$ is $L^p$. Thus $C = 0$.
Hence $\left| X_+\right| ^2=\left| X_-\right| ^2$ for $r>r_e$. The above
discussion shows that $\left| X_\pm \right|$ exists on $r=r_e$. So the match condition
of the solution gives $\left| X_+\right| =\left| X_-\right| =0$ on $r=r_e$. Therefore
$\Phi \equiv 0$ for $r \geq r_e$.

On $r_c <r<r_e$, $\Delta_{r}<0$ and $X_\pm(r)$ satisfy (\ref{radial2}), and on $0<r<r_c$,
$\Delta_{r}>0$ and $X_\pm(r)$ satisfy (\ref{radial1}). The same argument shows
that $\Phi \equiv 0$ for $r _c \leq r \leq r_e$ and $0 \leq r \leq r_c$. \qed

\begin{thm}\label{thm2}
Let $\Psi$ be the solution of (\ref{Dirac}) taking the form (\ref{phi}) on $r>0$ in non-extreme Kerr-Newman-AdS spacetime.
For $\frac{4}{3}<p\leq\frac{4}{3-2q}$, $0<q<\frac{3}{2}$, there is no nontrivial $\Psi$ for $|\lambda|>q \kappa $ which is $L^p $
over $M_{r_1}$ with some $r_1 >r_e$.
\end{thm}
\pf We only prove it for the case $\lambda>q \kappa$, and it can be proved in a similar manner for $\lambda<-q \kappa$.
By (\ref{radial1}), we obtain
\begin{eqnarray*}
\begin{aligned}
\frac{d}{dr}\left[i( \bar X_+ X_- - \bar X_- X_+)\right]
=& i\left[-i\alpha_1 \bar X_+ -(i\beta_1-\gamma_1)\bar X_-\right] X_- \\
&+ i\left[-i\alpha_1  X_- -(i\beta_1-\gamma_1) X_+\right]\bar X_+ \\
&- i\left[i\alpha_1 X_+ +(i\beta_1+\gamma_1)X_-\right]\bar X_- \\
&- i\left[i\alpha_1\bar  X_- +(i\beta_1+\gamma_1)\bar X _+\right]  X _+ \\
=&2\alpha_1 (\bar X_+ X_- + \bar X_- X_+)+2\beta_1 (|X_+ |^2 +| X_- |^2 )\\
\geq & 2(\beta _1 -|\alpha _1|)(|X_+ |^2 +| X_- |^2 ).
\end{aligned}
\end{eqnarray*}
As $\alpha_1\sim o(\frac{1}{r})$, $\beta_1\sim \frac{\lambda}{\kappa r}$ for sufficiently large $r$, there exists a sufficiently large $r_2 >r_1$
such that, for $r \geq r_2$,
\begin{eqnarray}
\frac{d}{dr}\left[i\left( \bar X_+ X_- - \bar X_- X_+\right)\right]\geq\frac{2q}{r}(|X_+ |^2 +| X_- |^2 ). \label{ineq}
\end{eqnarray}
We have already proved that
\begin{eqnarray*}
\left| X_+\right| ^2-\left | X_-\right |^2 = C
\end{eqnarray*}
for some constant $C$. If $C=0$, then $\Psi\equiv0$ from the proof of Theorem \ref{thm1}. Now we assume $C\neq0$, then (\ref{ineq}) gives
\begin{eqnarray*}
\frac{d}{dr}\left[i\left( \bar X_+ X_- - \bar X_- X_+\right)\right]\geq\frac{2q |C|}{r}
\end{eqnarray*}
for $r \geq r_2$. Integrating it, we obtain
\beQ
i\left[ \bar X_+ X_- - \bar X_- X_+\right]\geq 2q | C | \ln r +C'
\eeQ
for $r \geq r_2$. Thus there exists $r_3 >r_2$ such that, for $r \geq r_3$,
\beQ
i( \bar X_+ X_- - \bar X_- X_+)>0.
\eeQ
On the other hand, (\ref{ineq}) implies
\begin{eqnarray*}
\frac{d}{dr}\left[i\left( \bar X_+ X_- - \bar X_- X_+\right)\right] \geq \frac{2q}{r} i\left[ \bar X_+ X_- - \bar X_- X_+ \right]
\end{eqnarray*}
for $r \geq r_e$. Therefore, for $r \geq r_3$,
\beQ
\left| X \right| ^2 \geq i \left( \bar X_+ X_- - \bar X_- X_+ \right) \geq C_1 r^{2q}
\eeQ
for certain constant $C_1 >0$.
Thus, for $0<p\leq\frac{4}{3-2q}$,
\begin{eqnarray*}
\int_{M _{r_1} }\left |\Psi \right | ^p d\mu > C' \int_{r_3} ^\infty r^{1+(q-\frac{3}{2})p} dr =\infty
\end{eqnarray*}
This gives a contradiction. Hence $\Psi \equiv 0$ for $r \geq r_e$. Then, similar to the discussion in Theorem \ref{thm1}, we obtain
$\Psi \equiv 0$ for $r>0$ (Theorem \ref{thm1} already gives that $\Psi \equiv 0$ for $0<p\leq\frac{4}{3}$). \qed

\begin{coro}
By taking $q=\frac{1}{2}$ in Theorem \ref{thm2}, we obtain that there is no normalizable massive Dirac particle with mass greater than $\frac{\kappa}{2} $ outside and away from the event horizon in non-extreme Kerr-Newman-AdS spacetime, and they must either disappear into the black hole or escape to infinity.
\end{coro}

Note that this recovers the same result of Belgiorno and Cacciatori \cite{BC} in the case of $Q=0$ obtained by using spectral methods.

\begin{thm}\label{thm3}
Let $\Psi$ be the solution of (\ref{Dirac}) taking the form (\ref{phi}) on $r>0$ in non-extreme Kerr-Newman-AdS spacetime.
Any nontrivial $\Psi$ for $|\lambda| < \frac{\kappa}{2}$ must be $L^2$ over $M_{r_1}$ with some $r_1 >r_e$.
\end{thm}
\pf In terms of (\ref{radial1}), we have
\begin{eqnarray*}
\begin{aligned}
\frac{1}{2}\frac{d}{dr}\left(\left | X_+ \right | ^2 + \left | X_-\right | ^2\right)
=&\Re\left(\frac{d X_+}{dr} \bar X_+ \right)+\Re\left(\frac{d X_-}{dr} \bar X _- \right)\\
=&\Re\left(\left(i\alpha_1 X_+ +(i\beta_1+\gamma_1)X_-\right)\bar X_+ \right)\\
&+\Re\left(\left(-i\alpha_1 X_- +(-i\beta_1+\gamma_1)X _+\right) \bar X _-\right)\\
=&\Re\left(i\alpha_1 \left(X_+\bar X_+ - X_-\bar X_-\right)\right)\\
&+\Re\left(i\beta_1 \left( X_-\bar X_+ - X_+\bar X_-\right)\right)\\
&+\Re\left(\gamma_1 \left( X_-\bar X_+ + X_+\bar X_-\right)\right)\\
=&i\beta_1 \left( X_-\bar X_+ - X_+\bar X_-\right)+\gamma_1 \left( X_-\bar X_+ + X_+\bar X_-\right).
\end{aligned}
\end{eqnarray*}
As $\beta_1\sim \frac{\lambda}{\kappa r}$ and $\gamma_1\sim \frac{\epsilon}{\kappa r^2}$ for sufficiently large $r$, and
$ |\lambda |<\frac{\kappa }{2}$, there exists a sufficiently large $r_2 >r_1$ such that, for $r \geq r_2$,
\begin{eqnarray*}
\begin{aligned}
\frac{d}{dr}\left(\left | X_+ \right | ^2 + \left | X_-\right | ^2\right)\leq \frac{a}{r} (\left | X_+ \right | ^2 + \left | X_-\right | ^2)
\end{aligned}
\end{eqnarray*}
for some constant $0<a<1$. Therefore,
\begin{eqnarray*} \begin{aligned}
\left | X_+ \right | ^2 + \left | X_-\right | ^2\leq C r^a
\end{aligned}
\end{eqnarray*}
for some constant $C$.  Thus
\begin{eqnarray*}
\begin{aligned}
\int_{\{ r>r_1\}}\left |\Psi \right | ^2 d\mu
\leq C'\int_{r_1}^\infty r^{a-2}dr<\infty.
\end{aligned}
\end{eqnarray*}

\qed

Finally, we guess that there exists normalizable massive Dirac particle with mass less than $\frac{\kappa}{2}$ in non-extreme
Kerr-Newman-AdS spacetime. For the special case $a=0$ and spacetime is Schwarzschild-AdS:
\begin{eqnarray*}
\begin{aligned}
\widetilde{g}_{SAdS}=&-W d{t}^2+\frac{1}{W}d{r}^2 +r^2(d{\theta}^2+ \sin^2{\theta} d{\varphi}^2),
\end{aligned}
\end{eqnarray*}
where
\begin{eqnarray*}
\begin{aligned}
W&=1-\frac{2m}{r} +\kappa^2r^2=\frac{\Delta_{r}}{r^2},
\end{aligned}
\end{eqnarray*}
there does exist static, normalizable massless Dirac particle
\begin{equation*}
\Psi=e^{\frac{i}{2} \varphi}S^{-1}
\begin{pmatrix}
\sinh\big(-\int_{r_0}^rr^{-1} W^{-\frac{1}{2}}dr\big)\sin\frac{\theta}{2} \\
\cosh\big(-\int_{r_0}^rr^{-1} W^{-\frac{1}{2}}dr\big)\cos\frac{\theta}{2} \\
\cosh\big(-\int_{r_0}^rr^{-1} W^{-\frac{1}{2}}dr\big)\sin\frac{\theta}{2} \\
\sinh\big(-\int_{r_0}^rr^{-1} W^{-\frac{1}{2}}dr\big)\cos\frac{\theta}{2} \end{pmatrix}
\end{equation*}
with $k=-1$, $\varepsilon=-1$, $\omega=\lambda=0$.

\mysection{General stationary axisymmetric spacetimes}\ls

In this section, we use the argument in \cite{FKSY} to study the analogous nonexistence for general stationary axisymmetric spacetimes
with negative cosmological constant. The general stationary axisymmetric solutions for the Einstein field equations can be written as
\begin{eqnarray}\label{carter}
ds^2=\frac{\Delta_{\mu}(d\chi+\nu^2d\psi)^2- \Delta_{\nu}(d\chi-\mu^2d\psi)^2}{\nu^2+\mu^2}
+(\nu^2+\mu^2)\Big(\frac{d\nu^2}{\Delta_{\nu}} +\frac{d\mu^2}{\Delta_{\mu}}\Big),
\end{eqnarray}
with the potential
\begin{eqnarray}\label{carter1}
A=\frac{1}{\nu^2+\mu^2}\big(H(\nu)(d\chi-\mu^2d\psi)+K(\mu)(d\chi+\nu^2d\psi)\big),
\end{eqnarray}
where $\Delta_{\mu}$, $\Delta_{\nu}$ are smooth functions depending only on $\mu$, $\nu$ respectively,
$-\infty <\chi,\ \nu<\infty$, $0 \leq \psi < 2\pi$, $-a<\mu<a$ for some positive constant $a$.
We refer to \cite{C} for choices of these functions for Kerr-Newman-AdS spacetime.

Consider the Dirac equation
\begin{equation}\label{phi1}
\left(D+i\lambda \right)\Psi=0, \quad \Psi=S^{-1}\Phi, \quad
\Phi=e^{-i\left(\omega \chi+k\psi \right)}
\begin{pmatrix}
X_-(\nu)Y_-(\mu) \\
X_+(\nu)Y_+(\mu) \\
X_+(\nu)Y_-(\mu) \\
X_-(\nu)Y_+(\mu) \end{pmatrix}
\end{equation}
for real eigenvalue $\lambda$ and diagonal matrix
\begin{equation*}
S=\left|\Delta_{\nu}\right|^{\frac{1}{4}}
\mbox{diag}
\big((\nu+i\mu)^{\frac{1}{2}},
\big((\nu+i\mu)^{\frac{1}{2}},
\big((\nu-i\mu)^{\frac{1}{2}},
\big((\nu-i\mu)^{\frac{1}{2}}\big).
\end{equation*}
It is equivalent to the following equations in the region $\Delta_{\nu}>0$ (c.f. \cite{FKSY})
\begin{equation*}
\D _1 \phi=\D _2 \phi, \qquad
\phi=
\begin{pmatrix}
X_-(\nu)Y_-(\mu) \\
X_+(\nu)Y_+(\mu) \\
X_+(\nu)Y_-(\mu) \\
X_-(\nu)Y_+(\mu)
\end{pmatrix}
\end{equation*}
where
\begin{equation*}
\begin{aligned}
\D _1 & =
\begin{pmatrix}
-i\lambda \nu &  0 &  \sqrt{\Delta_{\nu}}D_{\nu+} & 0 \\
0   &     i\lambda \nu   &  0  &  \sqrt{\Delta_{\nu}}D_{\nu-}\\
\sqrt{\Delta_{\nu}}D_{\nu-}  &  0  & i\lambda \nu   &  0  \\
0 &  \sqrt{\Delta_{\nu}}D_{\nu+} &  0   &     -i\lambda \nu\end{pmatrix},\\
\D _2 &=
\begin{pmatrix}
\lambda\mu &  0 & 0 &  \sqrt{\Delta_{\mu}}L_{\mu+} \\
0   &     -\lambda\mu  & -\sqrt{\Delta_{\mu}}L_{\mu-} & 0 \\
 0  &  \sqrt{\Delta_{\mu}}L_{\mu+}  & \lambda\mu   &  0  \\
-\sqrt{\Delta_{\mu}}L_{\mu-} & 0 & 0   &     -\lambda\mu
\end{pmatrix}
\end{aligned}
\end{equation*}
with
\begin{eqnarray*}
\begin{aligned}
D_{\nu\pm}=&\frac{\partial}{\partial \nu}\mp\frac{i}{\Delta_{\nu}}\left(\omega \nu^2+H(\nu)-k\right).
\end{aligned}
\end{eqnarray*}
Thus there exists a real constant $\epsilon$ such that
\begin{eqnarray}\label{radial}
\begin{aligned}
\frac{d X_+(\nu)}{d\nu}-i\alpha X_+(\nu)-(i\beta+\gamma)X_-(\nu)&=0,\\
\frac{d X_-(\nu)}{d\nu}+i\alpha X_-(\nu)+(i\beta-\gamma)X_+(\nu)&=0
\end{aligned}
\end{eqnarray}
where
\begin{eqnarray*}
\begin{aligned}
\alpha=&\frac{1}{\Delta_{\nu}}\Big(\omega \nu^2+H(\nu)-k\Big),\ \
\beta=\frac{\lambda \nu}{\sqrt{\Delta_{\nu}}},\ \ \gamma=\frac{\epsilon}{\sqrt{\Delta_{\nu}}}.
\end{aligned}
\end{eqnarray*}
Similar equations can also be obtained in the region $\Delta_{\nu}<0$. Now we introduce the following conditions:
\begin{itemize}
\item[(A)] the asymptotic behavior $0<lim_{\nu\rightarrow\pm\infty}(\Delta_{\mu}-\nu^{-4}\mu^4\Delta_{\nu})<\infty$ since the the slice
$\{\chi=constant\}$ is asymptotically AdS;
\item[(B)] $\Delta_{\nu}$ has only simple roots $\nu_1<\cdots<\nu_e$, which are the horizons;
\item[(C)] $X_{+}=0$ or $X_{-}=0$ on the horizons which can match the solution inside and outside the horizons.
\end{itemize}

\begin{thm}\label{thm3}
Denote $M_{\nu_*}$ the slice $\{\chi=constant\}$ of the spacetime with $\nu \geq \nu_* >\nu_e$ for some $\nu_*>0$. Under the conditions $(A)$, $(B)$ and $(C)$, any solution $\Psi$ of (\ref{phi1}) which is $L^p$ over $M_{\nu_*}$ for $0<p\leq\frac{4}{3}$ must be zero.
\end{thm}
\pf In terms of (\ref{radial}) and $(B)$, we obtain: (i) $|\Phi |<\infty$ on $\nu=\nu_e$; (ii) $|\Phi |=0$ on $\nu=\nu_e$ implies that $|\Phi |=0$ on $\nu_e <\nu \leq \nu_*$.

On the other hand, (\ref{radial}) gives that
\begin{eqnarray*}
\begin{aligned}
\frac{d}{d\nu}\left(\left | X_+ \right | ^2 - \left | X_-\right | ^2\right)=0.
\end{aligned}
\end{eqnarray*}
This implies that
\begin{eqnarray*}
\begin{aligned}
\left| X_+\right| ^2-\left | X_-\right |^2 = C
\end{aligned}
\end{eqnarray*}
for some constant $C$. If $C\neq0$, we can assume $C > 0$ without loss of generality. Thus
\begin{eqnarray*}
\begin{aligned}
\int_{M_{\nu_*}}\left |\Psi \right | ^p dv
>C'\int_{\nu_*}^\infty \nu^{1-\frac{3p}{2}}d\nu =\infty
\end{aligned}
\end{eqnarray*}
for $0<p\leq\frac{4}{3}$ under the condition $(A)$. This contradicts the assumption that $\Psi$ is $L^p$. Thus $C = 0$.
Hence $\left| X_+\right| ^2=\left| X_-\right| ^2$ for $\nu>\nu_e$. The above discussion shows that $\left| X_\pm \right|$
exists on $\nu=\nu_e$. So the condition $(C)$ gives $\left| X_+\right| =\left| X_-\right| =0$ on $\nu=\nu_e$. Therefore
$\Phi \equiv 0$ for $\nu \geq \nu_e$. The similar argument can show that $\Phi \equiv 0$ on the whole slice. \qed

\mysection{Conclusion and future work}\ls

For non-extreme Kerr-Newman-AdS spacetimes, we proved that there is no nontrivial Dirac particle which is $L^p$ for $0<p\leq\frac{4}{3}$ with arbitrary eigenvalue $\lambda$, and for $\frac{4}{3}<p\leq\frac{4}{3-2q}$, $0<q<\frac{3}{2}$ with eigenvalue $|\lambda|>q \kappa $, outside and away from the event horizon. We also proved that any Dirac particle with eigenvalue $|\lambda|<\frac{\kappa}{2} $ must be $L^2$ outside and away from the event horizon. It concludes that there is no normalizable time-period Dirac particle with mass greater than $\frac{ \kappa}{2}$. In the future, we shall study whether there exists nontrivial normalizable time-period Dirac particle with mass less than or equal to $\frac{\kappa}{2} $?

\bigskip

{\footnotesize {\it Acknowledgement.}
This work is partially supported by the National Science Foundation of China (grants 11171328, 11571345, 11401168) and the project of mathematics and interdisciplinary sciences of Chinese Academy of Sciences. The authors would like to thank the referee for pointing out the paper of Belgiorno and Cacciatori as well as some valuable suggestions.

}

\end{document}